\documentclass[article,usenatbib]{mnras}

\usepackage{graphicx}
\usepackage{bm}
\usepackage{amsfonts}
\usepackage{latexsym}
\usepackage[latin1]{inputenc}
\usepackage{amsmath}
\usepackage{epsfig}
\usepackage{amsbsy}
\usepackage{psfrag}
\usepackage{mnfixes}

\usepackage{color}

\usepackage{bm}
\def\spose#1{\hbox to 0pt{#1\hss}}

\def\lta{\mathrel{\spose{\lower 3pt\hbox{$\mathchar"218$}}
     \raise 2.0pt\hbox{$\mathchar"13C$}}}
\def\gta{\mathrel{\spose{\lower 3pt\hbox{$\mathchar"218$}}
     \raise 2.0pt\hbox{$\mathchar"13E$}}}

\newcommand{\gsim}{\gtrsim}

\def\beq{\begin{equation}}
\def\eeq{\end{equation}}
\def\bea{\begin{eqnarray}}
\def\eea{\end{eqnarray}}

\usepackage{soul} 
\setstcolor{blue} 

\begin{document}

\title[Dynamical neutron star tides]{The phenomenology of dynamical neutron star tides}

\author[N. Andersson and P. Pnigouras]
{N. Andersson\thanks{E-mail:na@maths.soton.ac.uk} and P. Pnigouras \\ \\
School of Mathematics and STAG Research Centre, University of Southampton, Southampton SO17 1BJ, UK}

\maketitle

\date{\today}

\begin{abstract}
We introduce a  phenomenological, physically motivated, model for the effective tidal deformability of a neutron star, adding the frequency dependence (associated with the star's fundamental mode of oscillation) that comes into play during the late stages of the binary inspiral. Testing the model against alternative descriptions, we demonstrate that it provides an accurate representation of the dynamical tide up to close to merger. The simplicity of the prescription makes it an attractive alternative for a gravitational-wave data analysis implementation, facilitating an inexpensive construction of a large number of templates covering the relevant parameter space. 
\end{abstract}

\begin{keywords}
stars: neutron, neutron star mergers, gravitational waves\end{keywords}

\section{Introduction}

The inspiral and merger of binary neutron stars has long been considered a bread-and-butter source for advanced gravitational-wave detectors. Hence, the release of pent-up excitement following the spectacular GW170817 event \citep{ligo1,ligo2} came as no surprise. After all, these events have the potential to unlock scientific mysteries from astrophysics (confirming binary mergers as the origin of short gamma-ray bursts and explaining the cosmic generation of heavy elements) and cosmology (through inferred values of the Hubble constant) through to nuclear physics (as the imprint of matter on the gravitational-wave signal helps constrain the equation of state relevant for the extreme conditions represented by neutron stars). Remarkably, the  GW170817 event led to progress in all these directions \citep{grb1,grb2, kilonova,hubble,love} and the number of relevant analyses and discussions is already overwhelming.

The analysis of the GW170817 data has led to a---perhaps surprisingly tight---constraint on the neutron star tidal deformability, commonly expressed in terms of the dimensionless parameter  \citep{hind,hind0}
\beq
\Lambda_l =  {2\over (2l-1)!!} {k_l \over \mathcal C^{2l+1}} \ , 
\label{lamdef}
\eeq
where $k_l$ is the  Love number, $l$ is the relevant multipole (with the main contribution from the quadrupole, $l=2$) and $\mathcal C = GM_\star/Rc^2$ is the compactness of the star ($M_\star$ is the mass and $R$ the radius, and in the following we will use geometric units with $c=G=1$). Given that the  tidal imprint enters, formally, at the 5th post-Newtonian order \citep{hind}, the mass of the two binary partners can be inferred from lower order post-Newtonian terms in the signal. A constraint on $\Lambda_l$ can then be turned into a constraint on the neutron star radius \citep{love}. For example, the analysis of \citet{radice} suggests $400 \lesssim \tilde \Lambda_2 \lesssim 800$ (for a suitably averaged quadrupole tidal deformability $\tilde \Lambda_2$, depending on the mass ratio and with the lower limit somewhat model dependent). {This then leads to the radius being constrained to (roughly) the range 10-13~km, which overlaps with the first results for PSR J0030+0451 from NICER \citep{riley,colem}. The bottomline is that our current understanding of the neutron star radius comes with error bars at the 10\% level. This demonstrates how observations are beginning to constrain the theory and the neutron star equation of state, but we have not yet reached the (better than) 5\% error bars on the radius, the rough target for NICER and the point at which astrophysics observations would be doing ``better'' than current and upcoming nuclear physics experiments. Relevant efforts to constrain the strong interaction in laboratory experiments include the PREX effort, which probes the neutron skin thickness in lead, where a thick neutron skin would correspond to a large value for the neutron star tidal deformability
\citep{fatt}.}

The error bars on the neutron star radius will tighten with future NICER data and precision measurements of $\Lambda_2$ for neutron star binaries. However, the general expectation is that---unless we get very lucky!---the latter may require third-generation gravitational-wave instruments (like the Einstein Telescope or the Cosmic Explorer \citep{hall} or, perhaps, a dedicated high-frequency detector \citep{Ackley}), which might be able to constrain $\Lambda_2$ to the few percent level \citep{3G}. This should then, assuming that the neutron star masses can be inferred to higher precision, lead to a neutron star radius estimate accurate to the few percent level, as well.
This possibility motivates the discussion---the main point of which is to reduce the physics ambiguities and explore possible parameter degeneracies---in the present paper.

We outline a  phenomenological (yet, physically motivated) model for the dynamical tide in a neutron star binary, adding frequency dependence to the tidal deformability that comes into play during the late stages of inspiral. {This is a small effect, but it should be within reach of precision measurements for particularly ``bright'' future events. Moreover, the analysis is useful as it provides insight into the ``systematic errors'' associated with the tidal deformability, in the same way that our recent work \citep{ap1} shed light on the level at which the matter composition in the neutron star core comes into play. Moreover, the simplicity of the model makes it an attractive alternative for a data analysis implementation, which requires inexpensive construction of a large number of templates covering the relevant parameter space. Having said that, the aim of our discussion is manifestly not to develop such a model. Our focus is on the physics of the problem.}

\section{A simple phenomenological model}

We take as our starting point the discussion of \citet{ap1}, where the tidal response of a star is expressed in terms of the star's normal modes of oscillation\footnote{{As a slight aside, it is worth noting the similarity between the mode-sum for the tidal response and the sum-rule for the electric and magnetic dipole polarizability in atoms and nuclei \citep{Bernab, Mitroy}, yet another connection between  the astrophysics and laboratory experiments.}}. The original analysis  aimed to provide an idea of the ``error'' associated with the assumption that a deformed neutron star is described by a barotropic (beta equilibrium)  matter model rather than a model in which the matter composition is frozen as the system spirals through the sensitivity band of a gravitational-wave detector (which is expected as the timescale associated with nuclear reactions is much longer than that of the inspiral). {The question arises as most work on the tidal problem draws on phenomenological equation-of-state models (like piecewise polytropes or parameterisations based on the speed of sound), which do not account for fine-print issues like the state and composition of matter.}  The results from \citet{ap1} demonstrate that the dynamical contribution  to the tide is dominated by the excitation of the fundamental mode (f-mode) of the star. This is not surprising. The result was established a long time ago \citep{l94,reis,ks} in work aimed at quantifying the impact of mode resonances on the gravitational-wave signal (see \citet{ah} for a recent discussion of this problem), but the discussion from  \citet{ap1} adds a twist to the story. The results demonstrate that the sum over modes converges to the usual Love number in the static limit.  Again, this result could have been anticipated. As long as the modes form a complete set, they  can be used as a basis to describe any dynamical response of the star. This is well known, but the implications appear not to have been explored in previous work.

Let us make pragmatic use of the results from \citet{ap1} in order build a simple model that accounts for the main aspects of the dynamical tide. 
The basic idea is to include only the f-mode contribution to the mode sum and accept the contribution from other modes as a ``systematic error''.
Based on the stratified Newtonian models considered by \citet{ap1} we expect this error to be below the 5\% level. This level of uncertainty is smaller than our ignorance of (say) the neutron star equation of state, so the relation we write down should
 be precise enough for a ``practical'' construction of gravitational-wave templates. 

In essence, we start from a parameterised version of the Newtonian result for the effective Love number \citep{ap1}
\beq
k_l^\mathrm{eff}=   -{1\over 2} +{ A_f \over \tilde \omega_f^2 -\tilde \omega^2 }   \left( 1 - \tilde \omega^2 B_f\right) \left( 1 - \tilde \omega_f^2B_f\right)^{-1} \ , 
\label{keff1}\eeq
where $A_f$ depends on the overlap integral between the f-mode and the tidal driving, while $B_f$ involves the ratio of the horizontal and radial mode eigenfunctions at the star's surface. The f-mode frequency $\tilde \omega_f$ and the frequency associated with the Fourier transform (see \citet{ap1} for discussion) $\tilde \omega$ are both scaled to $\sqrt{GM/R^3}$. {We now insist that the relation \eqref{keff1} returns the usual Love number, $k_l$, in the static, $\tilde \omega\to0$, limit\footnote{This is the point where we ignore the formal contributions from other oscillations modes and, hence, the impact of the matter composition, which was the main focus of the discussion  of \citet{ap1}.}.  This means that we must have 
\beq
A_f    \left( 1 - \tilde \omega_f^2 B_f\right)^{-1} = \tilde \omega_f^2 \left( k_l + {1\over 2}\right) \ .
\label{coef}
\eeq
We can use this relation to replace one of the unknown parameters, $A_f$ or $B_f$. Opting to replace the former, we write \eqref{keff1} as
\begin{equation}
    k_l^\mathrm{eff} = {\tilde \omega_f^2 k_l \over \tilde \omega_f^2 - \tilde \omega^2} + {\tilde \omega^2 \over \tilde \omega_f^2  - \tilde \omega^2}  \left[ {1\over 2} - \tilde \omega_f^2 B_f \left( k_l + {1\over 2}\right)\right]
    \label{kleff1b}
\end{equation}
This expression is instructive. First of all, let us make the connection with an inspiralling binary by adding the usual assumption of adiabaticity, which links the Fourier frequency of the tidal response to the orbital frequency $\Omega$.  Focussing on the quadrupole modes, which make the main contribution to the gravitational-wave signal, we then have $\tilde \omega = 2\tilde \Omega$. We can also connect to the post-Newtonian expansion, which is commonly expressed in terms of  the dimensionless parameter $x=(\Omega M)^{2/3}$, where $M$ is the total mass of the system ($x=v^2$ where $v$ is the orbital velocity, and  the static tide enters the post-Newtonian expansion at order $v^5=x^{5/2}$, representing a 5th order contribution in the usual counting). From \eqref{kleff1b} we then see that---in contrast to other models that aim to account for the dynamical tide by including the main mode resonance (essentially adding a harmonic oscillator term to the adiabatic inspiral Hamiltonian), e.g. \citet{hind1} and \citet{stein}---the effective Love number is not (overall) proportional to the static one. However, the difference appears at a higher post-Newtonian order. Specifically, the difference enters as we, formally, add an order $x^{11/2}=v^{11}$ term to the expansion. The fact that this is a very high order contribution, which one would normally safely neglect, accords with the recent results of \citet{schmidt} (which, in turn, draw on the formulation from \citet{hind}). However, the argument is somewhat misleading because a formal low-frequency expansion fails to represent the f-mode resonance feature which (for typical neutron star parameters) will be prominent as the system approaches merger. This is, indeed, evident from the discussions of \citet{hind1} and \citet{stein}. A reasonable model has to retain the resonance feature. With this in mind, it is useful to note the result for incompressible stars \citep{ap1}. In this case, the relation 
\begin{equation}
    k_l^\mathrm{eff} =   {\tilde \omega_f^2 k_l \over \tilde \omega_f^2 - \tilde \omega^2}
    \label{kleff1c}
\end{equation}
is exact. That is, the leading term in the expression for the dynamical tide \eqref{kleff1b} takes the same form regardless of the matter compressibility. 
}

{Newtonian results---like \eqref{kleff1b}---lead to useful intuition for the qualitative behaviour, but they do not provide quantitative solutions to the actual problem (which obviously involves realistic neutron star models that require us to account for general relativity) we are interested in. However, the derivation of analogous results in the relativistic case pose a technical challenge and there has (at least so far) been very little work in this direction. This is unfortunate as it means that we do not yet have a precise model that can be  meaningfully compared to, for example, the  results of numerical simulations for the late stages of binary inspiral. However, if we assume that the form of the expression for the effective tide remains unchanged---and allow ourselves the freedom to introduce a couple of adjustable parameters---it is straightforward to write down a model that brings us closer to the result we need. 
}

{A logical first step in this direction involves the known ``universal relation'' between the f-mode frequency and the tidal deformability (see   \citet{chan}, as well as \citet{ak98,ilq}). This relation allows us to replace the mode frequency $\tilde \omega_f$ in \eqref{kleff1b} (say) with an expression involving $k_l$ (or, equivalently, $\Lambda_l$). This step should be ``safe'' given that the universal relation from \citet{chan} is robust\footnote{While we are confident in this argument, one should be aware that the universal relation involves a conservative view of the equation of state, e.g. the absence of sharp phase transitions \citep{han}. It would be useful to establish what happens for models that include phase transitions. However, if the relation breaks then so do related assumptions, like the I-Love-Q relations \citep{ilq} that are already used to break degeneracies in binary inspiral waveforms,  but there is no indication that this is the case \citep{pascha}. In that situation, one might still be able to make progress by separately constraining the tidal deformability and the f-mode frequency \citep{prat}.} (the evidence suggests that the errors involved are smaller than the error we introduce by assuming that the full mode-sum for the tide is replaced by the single contribution from the f-mode in the first place).
The relation we need takes the form
\begin{equation}
    \bar{\omega}_f = {a}_0 + {a}_1 y + {a}_2 y^2 + {a}_3 y^3 + {a}_4 y^4 \ , 
\label{omfit}
\end{equation}
where 
\begin{equation}
\bar{\omega}_f \equiv M_\star {\omega}_f =\tilde{\omega}_f\mathcal{C}^{3/2}
\end{equation} and $y = \ln \Lambda_l$ (with the $a_i$ parameters listed in Table~I of \citet{chan}). We see that we now need the stellar compactness $\mathcal C$. However, we can make use of another ``universal'' relation, this time between the compactness and the (quadrupole) tidal deformability \citep{mass}
\beq
\mathcal C \approx 3.71\times10^{-1} - 3.91\times 10^{-2} y + 1.056\times 10^{-3} y^2 \ , 
\eeq
to arrive at a one-parameter expression for the combination of the coefficients on the left-hand side of \eqref{coef}. Noting that $B_f=1/l$ for a homogeneous stellar model we represent the remaining parameter by $\epsilon$, such that $B_f=\epsilon/l$ (and it is worth noting that $\epsilon \approx 0.9$ for the polytropic models considered by \citet{ap1}). } 

We now have an explicit analytic formula for the effective Love number in terms of the result in the static limit, $k_l$, the (dimensionless) frequency $\bar \omega=\omega M_\star$ and the (to some extent) free parameter $\epsilon$. Moreover, even though it was based on a Newtonian analysis, the result makes use of fully relativistic relations  for the static Love number and the mode frequency. {In order to make the connection with the gravitational-wave signal, we relate $\bar \omega$ to the (similarly scaled) orbital frequency $\bar \Omega$. Focussing  on the $l=m=2$ resonance we then have (for an equal-mass binary)
\begin{equation}
\bar \omega = 2 \bar \Omega = 2 \Omega M_\star = \Omega M = x^{3/2}
\end{equation}
where $M$ is the total mass of the system, as before, and $x$ is the post-Newtonian expansion parameter.
}

{Finally, we note that the mode frequency  in \eqref{omfit} includes the gravitational redshift (i.e. refers to an observer at infinity), while the tidal interaction involves a binary partner at a finite distance. In order to account for this we  introduce a second free parameter, $\delta$ , such $\bar \omega_f^2 \to \bar \omega_f^2/\delta$. This then leads to the expression
\beq
k_l^\mathrm{eff} \approx {\bar \omega_f^2 k_l \over \bar \omega_f^2 - \delta (2\bar \Omega)^2} +{(2\bar \Omega)^2 \over \bar \omega_f^2 - \delta (2\bar \Omega)^2}\left[ {\delta \over 2} - {\bar \omega_f^2 \over \mathcal C^3 } {\epsilon\over l} \left( k_l + {1\over 2} \right) \right]\ .
\label{kleff3}
\eeq
The new parameter $\delta$ has to account for two relativistic effects. First, we have the gravitational redshift of the mode frequencies that we have already alluded to. Secondly, we need to consider the rotational frame-dragging induced by the orbital motion. The problem is not trivial, but an argument by \citet{stein} suggests that the two effects almost cancel so---as a first approximation---we are motivated to  simply remove the gravitational redshift altogether. Simplistically, for a signal emitted at $r_0$ and observed at $r_1$ we have
\begin{equation}
{\omega_1 \over \omega_0} = \left( {1-2M_\star/r_0 \over  1-2M_\star/r_1} \right)^{1/2}.
\end{equation}
If $\omega_0$ is the frequency "emitted" at $R$ and $\omega_f$ is the frequency observed at infinity, then
\begin{equation}
\omega_f^2=\left(1-2\mathcal{C}\right)\omega_0^2.
\end{equation}
 Keeping in mind that the mode frequencies we use include the gravitational redshift, we may simply take $\delta = 1- 2 \mathcal C$  to remove it.
 This then reduces the result to a one-parameter expression for the effective Love number. As we will now demonstrate, this expression performs surprisingly well.
}

As a suitable comparison (in fact, the only comparable results in the literature), we consider the results
 for the dynamical tide from \citet{hind1} and \citet{stein},  which are similar in spirit as they introduce the notion of an effective tidal deformability. However, the main focus of \citet{hind1} and \citet{stein} was to extend the effective-one-body framework to account for the dynamical tide. In addition to this, \citet{stein} provides an approximate analytical formula from matched asymptotics. This result has been tested against numerical relativity simulations, see for example \citet{fouc}. As it appears to perform well in such comparisons \citep{diet1}, it provides a natural benchmark against which to test our simple closed-form expression. 

\begin{figure}
\begin{center}
\includegraphics[width=0.9\columnwidth]{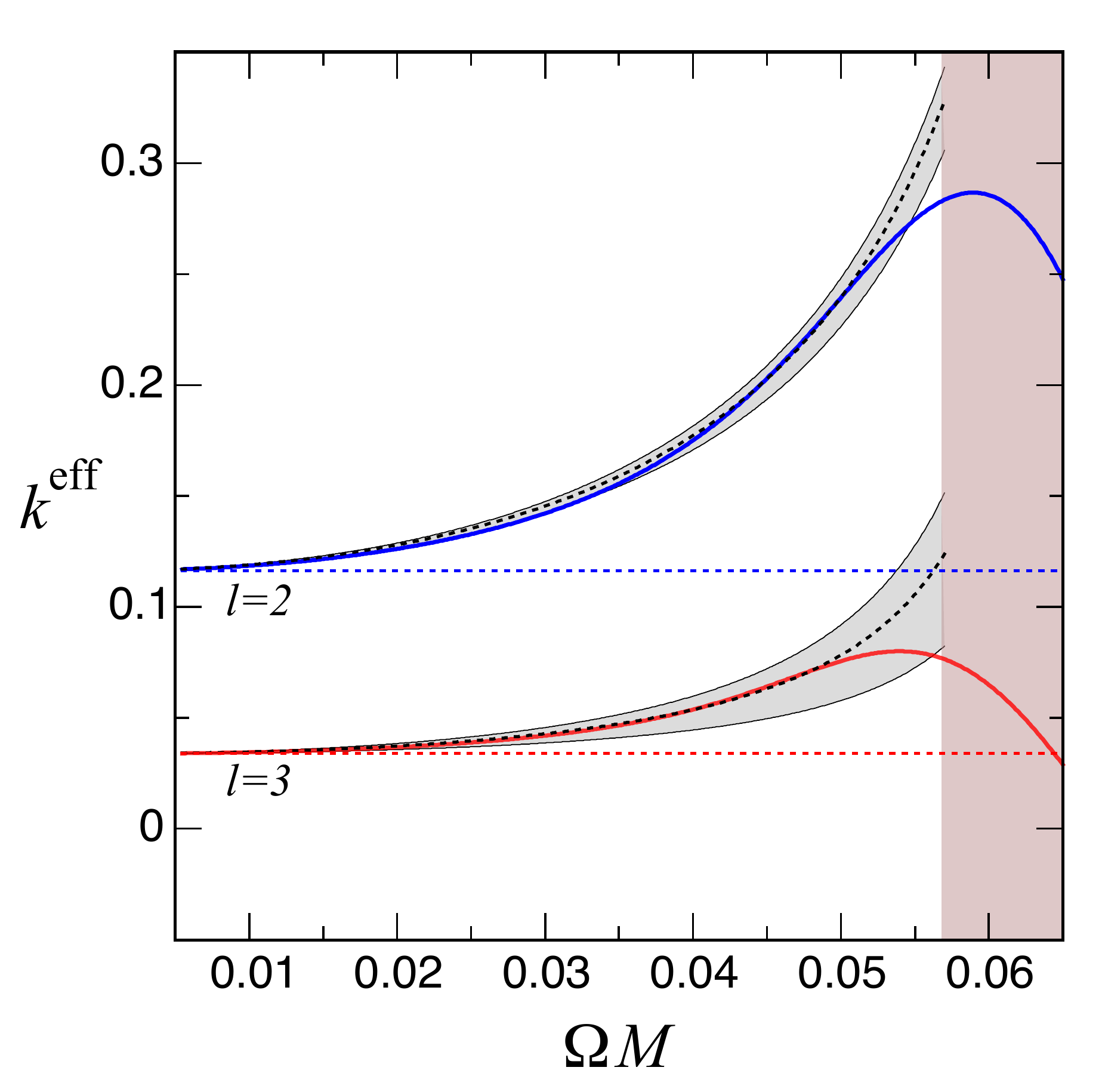}
\end{center}
\caption{Comparing the effective Love number $k_l^\mathrm{eff}$ from \eqref{kleff3} with $\delta = 1- 2 \mathcal C$ to the results from \citet{stein}, for both $l=2$ and $l=3$. The dashed horizontal lines represent the static Love number ($k_l^\mathrm{eff}$ in the $\Omega\to 0 $ limit). The results of \citet{stein} are shown as solid curves (blue for $l=2$ and red for $l=3$).  Estimates from \eqref{kleff3} are shown for the range $\epsilon=0.85-0.9$ with the latter representing the lower edge of the filled band in each case. The particular choice $\epsilon=0.875$ (dashed black curves) provides an an excellent fit to the results from \citet{stein}. Finally, we indicate the region beyond the (approximate) merger frequency, $\Omega M\gsim 0.057$ in this case, by the shaded area in the figure. } 
\label{ltest}
\end{figure}

Focussing on the example illustrated by \citet{stein}, i.e., an equal mass neutron star binary with $M_\star = 1.350 M_\odot$ and $R=$13.5~km, leading to $\mathcal C=0.148$ and $\Lambda_2 =1111$, it is 
easy to demonstrate that we obtain an accurate representation of the dynamical tide throughout the relevant frequency range (up to close to merger where our simple prescription obviously breaks as the expression for $k_l^\mathrm{eff}$ diverges\footnote{In principle, it is fairly straightforward to regularize the resonance should one want  to do so, see for example the stationary phase approximation used by \cite{stein}. However, as it is not clear how this procedure changes when the problem is considered in general relativity, and the spirit of our strategy is to facilitate an intuitive extension in that direction, we will leave the singular behaviour in the phenomenological model.}) by
tuning the parameter $\epsilon$. 
The effective Love number obtained from \eqref{kleff3} is compared to the results from \citet{stein} in figure~\ref{ltest}, for both $l=2$ and $l=3$. As our formula still leaves $\epsilon$ as a free parameter, we show results for the range $\epsilon=0.85-0.9$ and, as is clear from the result in the figure, the corresponding curves for $\epsilon=0.875$ provide an excellent match to the results from \citet{stein}. {It is also worth noting that $\epsilon\approx 0.935$ removes the second term in \eqref{kleff3}, leaving an expression of the form expected from incompressible models. We note from the figure that this would tend to underestimate the value for $k_l^\mathrm{eff}$ (the corresponding result would lie beneath the grey regions shown in the figure), which indicates the level at which the parameter $\epsilon$ impacts on the result. } The essence of the comparison is that our formula \eqref{kleff3} performs (perhaps surprisingly) well. It may be phenomenological in origin, but there can be little doubt that our simple expression provides an effective representation of the required behaviour, and hence reflects the underlying physics. 

It is worth noting that, while the two sets of results diverge for large values of $\Omega M$ in Figure~\ref{ltest}, the corresponding frequencies are close to (or indeed beyond) the merger frequency. As the picture of two separate, tidally deformed, bodies breaks down there is no reason to expect the model to make sense beyond this point. The post-merger region is indicated by the shaded area in Figure~\ref{ltest}. In order to estimate the merger frequency, we have simply taken the corresponding  orbital separation to be the sum of the neutron star radii, which leads to $\Omega M \approx\mathcal C^{3/2}$.
For the model used in Figure~\ref{ltest}  the estimated merger frequency would then be $\Omega M \approx 0.057$ (corresponding to a gravitational-wave frequency of about 1,400~Hz for this specific model). A more precise estimate could be obtained from the results of \citet{read}, but this would not change the conclusions.

\section{The state of play}

The favourable comparison to the results from \citet{stein} suggests that the simple relation for the effective tidal deformability \eqref{kleff3}  remains useful up to close to the final merger. The main lesson from this is that the phenomenology of the problem is clear. The discussion admittedly does not add much to the well established logic developed for the Newtonian tidal problem \citep{l94,reis,ks,ap1}, but the steps we have taken towards a relativistic model provide us with a better handle on the involved systematics. Moreover,  building on this, the evaluation of the large set of templates required to span the parameter space relevant for gravitational-wave searches should not be computationally costly. In particular, it would be straightforward to combine \eqref{kleff3} with any current waveform model that implements the static tide. However, the emphasis of this paper is not on the development of an alternative waveform model---there are many such efforts in the literature. Rather, we are interested in the physics of the tidal problem, to what extent we  can develop a simple representation for the dynamical tide  and---now that this has been demonstrated---to what extent such a model provides useful insight. One particular issue, which will inevitably come to the fore as the discussion of third-generation ground-based detectors gathers pace, relates to the precision with which the tidal information can be inferred from observations and turned into constraints on the neutron star radius and the matter equation of state. With this in mind, it is evident from Figure~\ref{ltest} that the dynamical tide dominates during the late stages of inspiral (as the system  approaches resonance). The simple fact that it strengthens the tidal imprint should be good for observations and, combined with the simple link to the static tide provided by (say) \eqref{kleff3}, one might hope to be able to facilitate a more precise extraction of the Love number $k_2$ and hence more secure radius constraints. A related aspect concerns the measurement uncertainties. Assuming that the observational errors of the tidal deformability $\Lambda_2$ may reach the percent level, how do we make sure that the parameter inference is not limited by the theory? Could it, for example,  be the case that this level of precision requires information beyond bulk properties like mass and radius (e.g. the internal composition discussion by \citet{ap1} or the presence of superfluidity in the neutron star core considered by \citet{wein})?

As we reflect on the different options, we need to consider the all-important benchmarking of phenomenological (computationally efficient) models against (computationally costly) nonlinear numerical simulations. This is relevant for many reasons. Perhaps most importantly, while an absolute requirement for a description of the complex merger dynamics, numerical simulations are unlikely to be able to track the many thousand binary orbits required to model a system that evolves through the detector sensitivity band. This will always require an approximate description. For well-separated binaries, this need is satisfied by post-Newtonian (essentially point particle) results but this description becomes less reliable during the late stages of inspiral---largely due to the  emergence of finite size effects, like the tidal deformability. The importance of the problem, both for signal detection and the extraction of parameters from an observation, has driven the development of reliable alternatives involving dynamical aspects of the tide, like the effective-one-body framework that forms the basis for the work of \citet{hind1} and \citet{stein}.  

An important point, which appears to be commonly overlooked {but is evident from the discussion leading to \eqref{kleff3}}, involves the natural parameters to use in a phenomenological model. The typical approach connects with post-Newtonian logic by expressing the results in terms of  the dimensionless parameter $x=(\Omega M)^{2/3}$.
This is an obvious choice for the main contribution to the gravitational-wave signal, which depends on the orbital motion, but it  is less clear that this parameter makes sense during the late stages of inspiral, for which numerical simulations are viable (recall that it is rare that binary neutron star simulations are carried out for more than the last 10-20 orbits and it is only very recently that such simulations have been carried out with sub-radian precision in the accumulated gravitational-wave phase \citep{diet2,fouc,kiuchi,kawa}). There is no reason why the matter effects would be naturally expressed in terms of a parameter based on the orbital dynamics in this regime. In fact, this would be counter-intuitive. This point is  neatly illustrated by \eqref{kleff3} which encodes the matter dynamics (for each binary companion) in terms of the f-mode oscillation frequency, a behaviour that would be obscured by a formal expansion in $x$.

\begin{figure}
\begin{center}
\includegraphics[width=0.9\columnwidth]{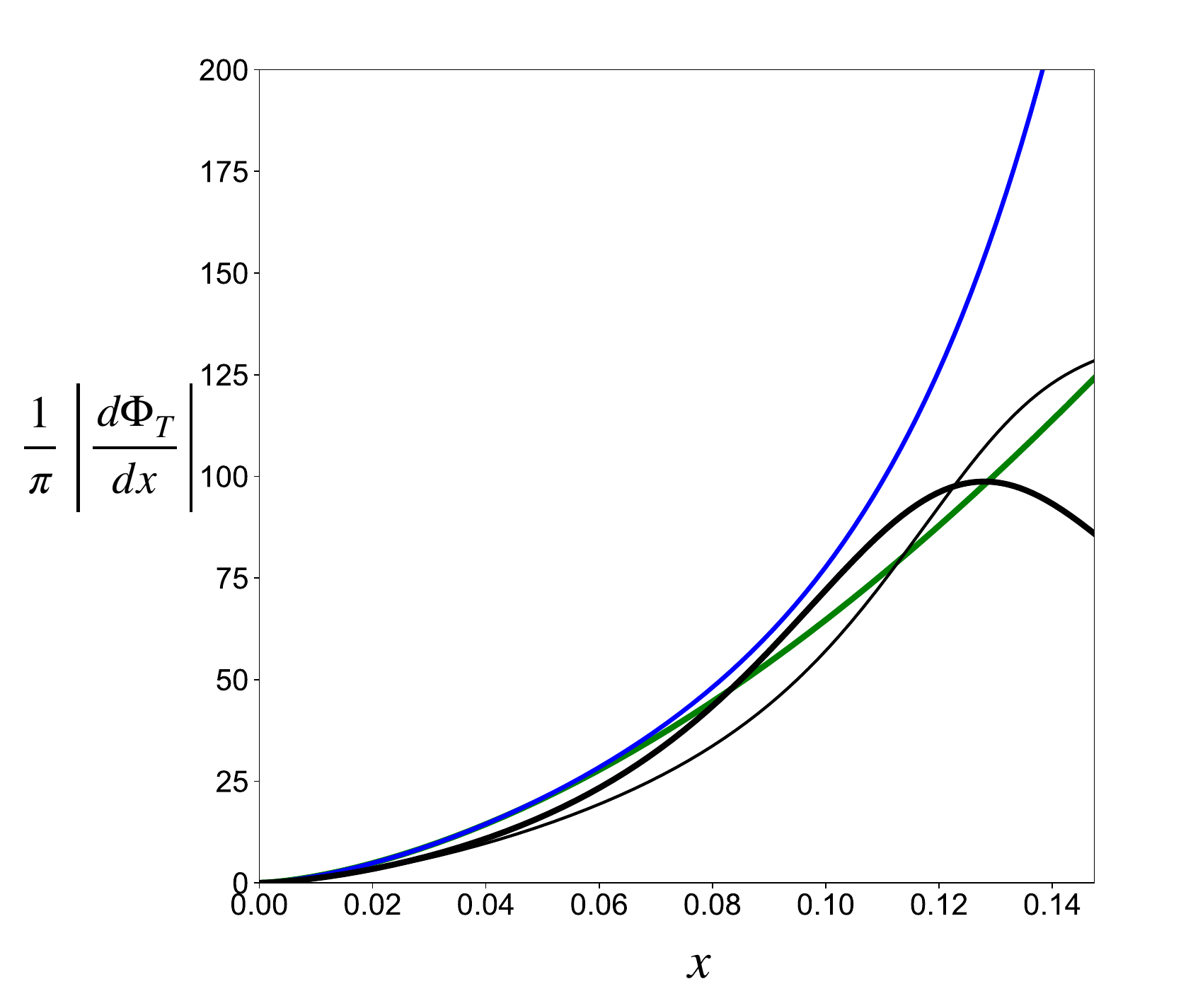}
\end{center}
\caption{Comparing the impact of the tidal deformability for different closed form expressions based on high-order post-Newtonian calculations or nonlinear inspiral simulations (all for $l=2$). We show the derivative (with respect to $x$) of the tidal contribution to the gravitational-wave phase.  From top to bottom on the right edge, the curves represent: The post-Newtonian model from \citet{damour} (solid blue curve), the fit to numerical data from \citet{diet2} (thin black curve), a fit to the numerical data from \citet{kawa} (solid green curve)  and the recent fit to numerical data from \citet{diet5} (thick  black curve). The results illustrate the level of ``uncertainty'' in  current state-of-the-art models.  } 
\label{numfig}
\end{figure}

As a  measure of the current level of uncertainty, let us compare different suggested models for the tidal contribution. This kind of comparison is straightforward, as several alternatives have been given in closed form. However, one has to be careful because the associated assumptions impact on the result. We need to compare apples with apples.  As will soon become clear, this turns out to be less straightforward.
A relevant comparison, with immediate implication for gravitational-wave astronomy, involves the accumulated phase associated with the tidal contribution. At the very least, one would expect to be able to distinguish between models that differ by at least half a cycle (one radian) in the inspiral signal {(although the large signal-to-noise detections expected from third generation detectors should allow much better precision than this)}. Effectively, for the quadrupole contribution to an equal mass binary signal we need to integrate an expression of  form \citep{hind0,ah}
\beq
{d\Phi_T \over dx} = - {65\over 2^5} {k_2 \over \mathcal C^5} x^{3/2} f(x)
\label{phder}\eeq
for the tidal contribution to the phase, $\Phi_T$. The Newtonian prefactor is the same for all models, but the factor $f(x)$ differs, {e.g. depending on which point-particle inpiral model we consider}. Since the models we consider can be described analytically, it is easy to obtain an idea of the difference between the tidal prescriptions. 

{Figure~\ref{numfig} provides a summary of the current state of the art (for the same stellar model as before). The figure shows (all for $l=2$): the post-Newtonian model from \citet{damour}, the fit to numerical data from \citet{diet2}, a similar fit from \citet{kawa} and the recent expression from \citet{diet5}. The last three of these models are based on (different) sets of nonlinear simulations for the late stages of inspiral, matched to a chosen post-Newtonian model for low frequencies. The results in Figure~\ref{numfig} show\footnote{We have treated the various expressions for the accumulated phase $\Phi_T$ as ``exact'' and taken the derivative. As the expressions we consider involve order by order post-Newtonian expansion or Pad\'e approximants to improve the models, this derivative is not particularly consistent, but the results we should should be considered as illustrative so this should not be too much of a concern.} that there is significant variation between the models, suggesting that the discussion of the problem has not yet ``converged''.
In principle, the discussion should implement as much ``known'' post-Newtonian information as possible. In practice, this involves deciding which post-Newtonian (or, indeed, effective-one-body) model one should take as the baseline. The problem is that one would expect post-Newtonian estimates to lead to larger neutron stars and hence an enhanced tidal effect \citep{diet3}. This expectation is illustrated by the results for the post-Newtonian model from \citet{damour}  (see their equation (31)), which differs dramatically  from fits based on numerical simulations for $x\gsim 0.1$ (see Figure~\ref{numfig}). 
Another key take-home message from the data concerns the low-frequency behaviour. We note that the approximation from \citet{kawa} asymptotes to the result from \cite{damour} as $x\to0$, while the low-frequency slope of the two models from \citet{diet2} and \citet{diet5}  (see also \citet{diet3,diet4}) is different. The latter limit to the integrated version of \eqref{phder} with a fixed $k_2$ and an $f(x)$ such that the relevant factor in the phase is  $1+c_1x$ with $c_1=3115/624$ (in agreement with  the post-Newtonian correction from \citet{vines}).
This observation relates directly to the fact that the dynamical tide arises at a high post-Newtonian order (see the analysis of \citet{schmidt}). In effect, we need to account for this before we compare an expression like \eqref{kleff3} to the existing models. Effectively, noting that the function $f(x)$ represents non-dynamical aspects of the tide, we need to decide which model to use as our benchmark for comparison. 
For natural reasons, as it is the most recent discussion of the problem, we will compare to the result from \citet{diet5} in the following. }

\begin{figure}
\begin{center}
\includegraphics[width=0.9\columnwidth]{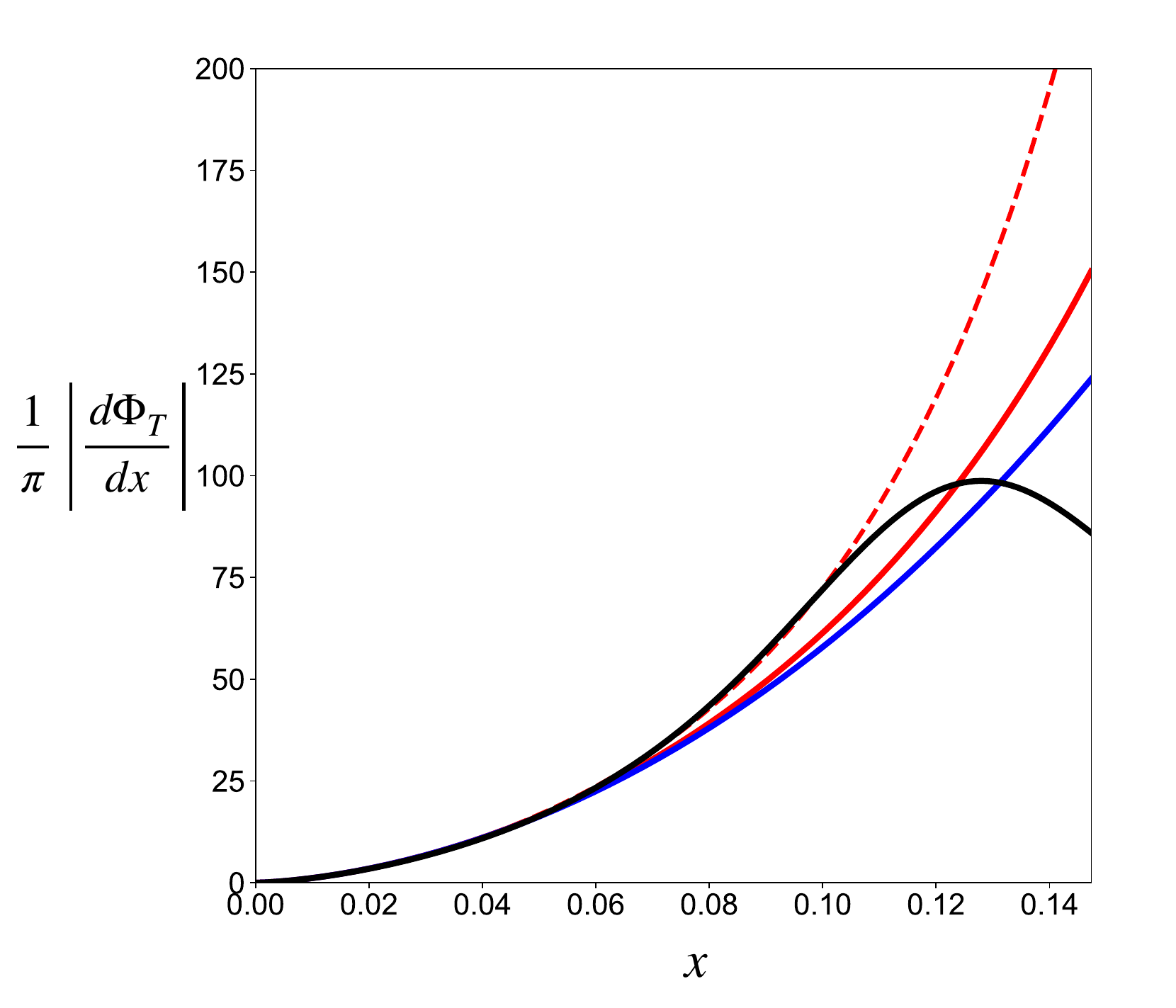}
\end{center}
\caption{Different approximate results for the tidal contribution, taking the recent results from \citet{diet5} (solid black curve, based on a selection of numerical relativity simulations) as the benchmark for comparison. In turn, we show the static contribution to the tide (as a blue curve), our approximate expression for the dynamical tide from \eqref{kleff3} (solid red curve, using the ``best-fit'' value $\epsilon\approx 0.875$, which we know (from Figure~\ref{ltest}) is close to the result from \citet{hind1} for much of the inspiral). We also show that simply taking $\epsilon=0$ (dashed red curve) in \eqref{kleff3} leads to a surprisingly good approximation to the numerical relativity result for much of the frequency range. } 
\label{appfig}
\end{figure}

\begin{figure}
\begin{center}
\includegraphics[width=0.9\columnwidth]{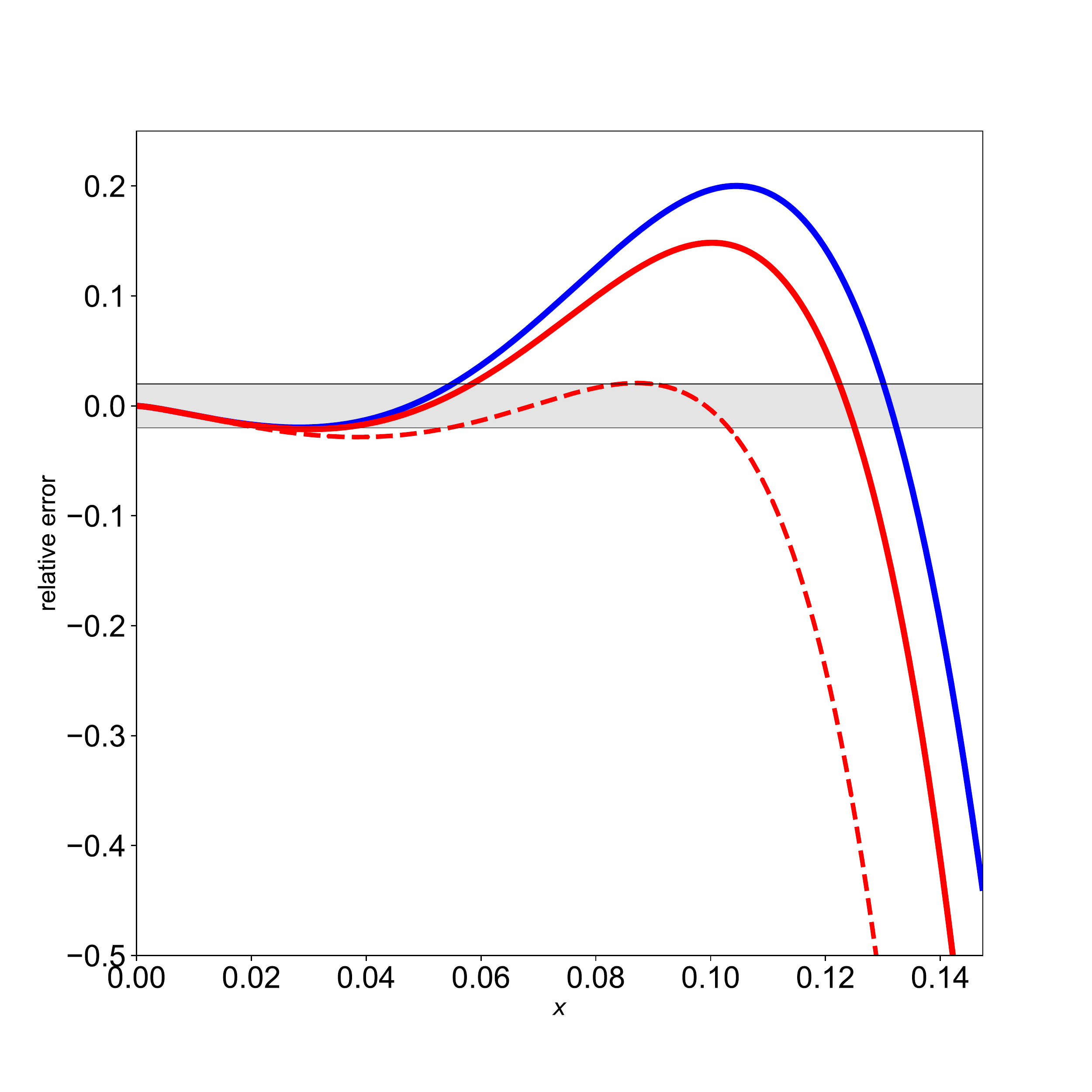}
\end{center}
\caption{Same as Figure~\ref{appfig}, but showing the relative difference between the different approximations and the numerical relativity based model from \citet{diet5}. As a rough indication of the accuracy of the different models for the tidal contribution, we show (as a grey horizontal band) the $\pm 2\%$ error band. } 
\label{relfig}
\end{figure}

{The results in Figure~\ref{appfig} provide a  comparison of the model based on \eqref{kleff3} and the result from \citet{diet5}. First of all, we can compare the impact of the effective tidal deformability from \eqref{kleff3} to the static tide. Figure~\ref{appfig} then indicates that the contribution from the dynamical tide brings us closer to the numerical relativity fit than the static tide. This is not a surprise---the numerical simulations of \citet{fouc} suggest an  agreement with the effective tidal formulation from \citet{stein} (and by implication from Figure~\ref{ltest}, our expression). However, in the case we consider here the match is not great beyond something like $x\approx 0.06$ (roughly corresponding to a gravitational-wave frequency of 350~Hz). It turns out that we can improve the match by making use of the free parameter ($\epsilon$). Somewhat surprisingly, the phenomenological model performs well up to significantly higher frequencies if we replace the
 ``best-fit'' value $\epsilon\approx 0.875$ suggested by Figure~\ref{ltest} with $\epsilon=0$. It is not clear, at least not at this point, why this should be the case---or, indeed, if this choice (which represent the horizontal component of the displacement vanishing at the surface) makes physical sense---but the model now works well up to about $x\approx 0.1$ (roughly 760~Hz). The result is emphasized by Figure~\ref{relfig}, which shows the relative difference between the model from \eqref{kleff3} and the expression from \citet{diet5}. In this case we also provide a 2\% error band as an indication of the level of precision that may be required when we consider third-generation detectors. The theory has clearly not reached this level yet.
}

\begin{figure}
\begin{center}
\includegraphics[width=0.9\columnwidth]{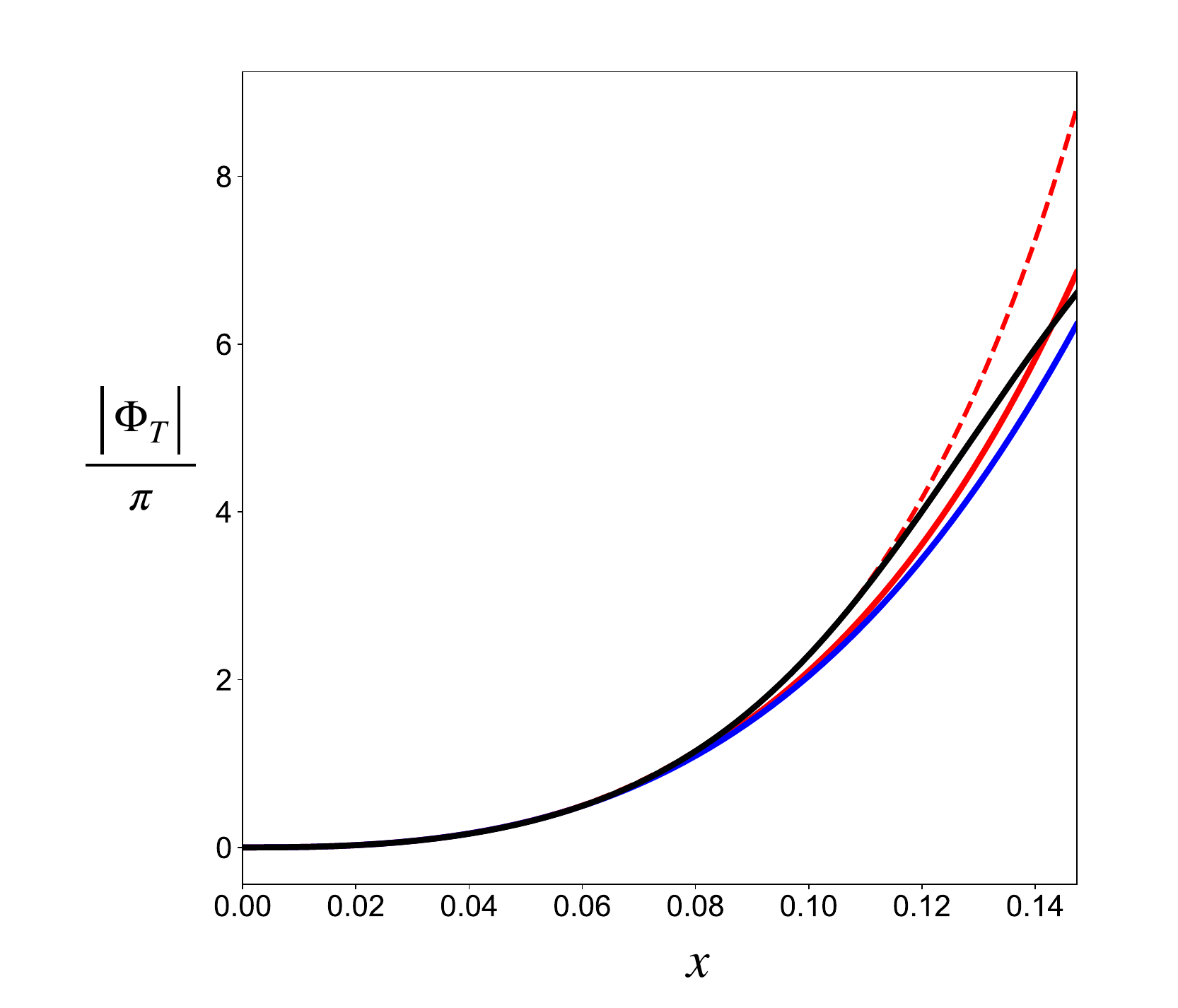}
\end{center}
\caption{The accumulated gravitational-wave phase $\Phi_T$ for the models considered in Figure~\ref{appfig}, highlighting the expected result that our simple expression \eqref{kleff3} extends the frequency range across which the approximation remains faithful to the results inferred from numerical relativity (in this case, the model from \citet{diet5}), but also that the result diverges for the very late stages of inspiral (as expected).} 
\label{phasefig}
\end{figure}

Finally, turning to the accumulated phase, illustrated (for the same set of models as in Figures~\ref{appfig} and \ref{relfig}) in Figure~\ref{phasefig}, we see that the effective dynamical tide leads to a slight (sub-radian) change in the gravitational-wave phasing. This is a small effect, but it should be distinguishable by the high signal-to-noise detections expected by third-generation detectors \citep{hall,3G} or dedicated high-frequency instruments \citep{Ackley}. 
In this case it is worth noting that the ``best fit'' value of $\epsilon\approx 0.875$ from Figure~\ref{ltest} performs much better than the ad hoc $\epsilon=0$ above $x\approx 0.12$, i.e. just before merger. This is simply a reflection of the fact that the phenomenological expression from \eqref{kleff3} diverges as we approach the mode resonance (see footnote 4). It is also worth noting that the same phenomenological model stays close to the numerical fit from \cite{diet2} provided we change $c_1\to c_1/2$. In fact, in that case the models agree well up to $x\approx 0.12$ (a gravitational-wave frequency close to 1,200~Hz).

{In summary, the phenomenological model from \eqref{kleff3} performs well in comparisons with models drawing on numerical simulations, and hence provides  useful understanding of the main physics that need to be accounted for in a model for the dynamical tide. }
 Of course, the issue of the ``correct'' form for $f(x)$ requires further thinking. The comparison with numerical relativity also requires some care as one would anticipate numerical dissipation to enhance the energy loss in the system, leading to a faster inspiral in nonlinear simulations and this may (to some extent) mimic the tide. It could be that the simulations do not yet have the level of precision we need for a true comparison. The results from \citet{fouc} would seem to support this. However, one has to be careful. It is notable that, even though the three numerical-relativity inspired descriptions in Figure~\ref{numfig} agree to sub-radian precision in the overall gravitational-wave phase, the formulas from \citet{kawa} and  \citet{diet2,diet5} match different models in the low-frequency limit. Keeping in mind that the numerical simulations involve only the final 10-20 orbits, one should really focus on the region above $x\approx 0.1$ in the different figures. This brings us to another key point, where further deliberation is needed. A given numerical simulation does not automatically represent the past history of a binary inspiral. The initial data does not have the required ``memory'' (it may, for example, involve some level of unwanted eccentricity \citep{diet1,bern}). As simulations become more precise (with differences at the sub-radian level required for the results to be reliable enough for gravitational-wave data analysis) the matching to the low-frequency part of the inspiral signal inevitably comes to the fore. A better understanding of the physics associated with the tidal response should be valuable for this effort.

\section{Final remarks}

We have introduced a  phenomenological, physically motivated, model for the effective tidal deformability of a neutron star binary, adding frequency dependence that comes into play during the late stages of inspiral. A comparison against alternative descriptions (like the results from \citet{stein}), suggests that we have at hand a simple, yet accurate, description of the tidal imprint. This should make the model an attractive alternative for an implementation of the matter effects in gravitational-wave data analysis algorithms. The remaining free parameter of the model ($\epsilon)$ tends not to have much impact on the gravitational-wave phasing, unless we push it outside the range expected from the results of \cite{ap1}. Having said that, we have noted that simply setting this parameter to zero (leaving us with no free parameters, at all!) leads to a model that agrees very well with expressions for the dynamical tide based on fits to numerical simulations (see, in particular, Figure~\ref{relfig}). It is not clear  at this point why this should be the case; the issue requires further exploration. 

Our results suggest interesting avenues for future work.
For example, it ought to be straightforward to extend the logic to rotating systems by making use of the phenomenological relations from \citet{doneva} which encode the effect that spin has on the fundamental mode (although it is worth keeping in mind that merging
neutron star binaries are likely to be old enough that the stars will have had plenty of time
to spin down, {so an assumption of slow spin may well be adequate}).
At a more formal (and challenging) level, we need to extend the mode-sum approach from \citet{ap1} to general relativity. This is an essential step if we want to base the analysis on the use of realistic matter equations of state. Efforts in this direction should  allow an actual derivation of a result along the lines of \eqref{kleff3} (rather than the present argument, which was based on analogy with the Newtonian analysis). However, the relativistic problem is technically difficult because the stellar oscillation modes are quasi-normal (with inevitable damping due to gravitational-wave emission) and  known not to be complete (as the scattering of waves by the spacetime curvature leads to a late-time power-law tail \citep{gund}). Nevertheless, this should be a priority issue, as one would also arrive at a precise description  of mode resonances \citep{ah}. The importance of such a development is clear, but the technical challenge should not be underestimated. 

\section*{Acknowledgments}
	We would like to thank Sebastiano Bernuzzi for helpful conversations. Support from STFC via grant ST/R00045X/1 is gratefully acknowledged.

\section*{Data availability}

All relevant data required to reproduce the results are incorporated into the article. Additional material available on request.

\end{document}